\documentclass[12pt,preprint]{aastex}

\newcommand{\onee}  {\mbox{\rm\,1E~1740.7--2942}}
\newcommand{\lmc}  {\mbox{\rm\,LMC~X--3}}
\newcommand{\lmo}  {\mbox{\rm\,LMC~X--1}}
\newcommand{\cyg}  {\mbox{\rm\,Cyg~X--1}}
\newcommand{\gx}  {\mbox{\rm\,GX~339--4}}

\newcommand{\grs}  {\mbox{\rm\,GRS~1758--258}}

\newcommand{\mdotd}     {$\dot{M}_{\rm disk}$}
\newcommand{\mdotc}     {$\dot{M}_{\rm corona}$}

\begin{document}

\title{Hysteresis of spectral evolution in the soft state of black-hole
binary LMC~X--3}

\author{D. M. Smith\altaffilmark{1}, D. M. Dawson\altaffilmark{1},
and J. H. Swank\altaffilmark{2}}

\altaffiltext{1}{Physics Department and Santa Cruz Institute for Particle Physics,
University of California, Santa Cruz, 1156 High St., Santa Cruz, CA 95064}
\altaffiltext{2}{NASA's Goddard Space Flight Center, Code 662, Greenbelt, MD, 20771}

\begin{sloppypar}

\begin{abstract}

We report the discovery of hysteresis between the x-ray spectrum and
luminosity of black-hole binary \lmc.  Our observations, with the
Proportional Counter Array on the Rossi X-ray Timing Explorer, took
place entirely within the soft spectral state, dominated by a spectral
component that was fitted well with a multicolor disk blackbody.  A
power-law component was seen only during times when the luminosity of
the disk blackbody was declining. The x-ray luminosity at these times
was comparable to that seen in transient systems (x-ray novae) when
they return to the hard state at the end of an outburst.  Our
observations may represent partial transitions to the hard state;
complete transitions have been seen in this system by \cite{Wi01}. If
they are related to the soft-to-hard transition in transients, then
they demonstrate that hysteresis effects can appear without a full
state transition.  We discuss these observations in the context of
earlier observations of hysteresis within the hard state of binaries
\onee\ and \grs\ and in relation to published explanations of
hysteresis in transients.

\end{abstract}


\section{Introduction}

There are five persistently active black-hole-binary candidates:
\lmc, \lmo, \cyg, \onee\ and \grs.  The latter two are candidates by
courtesy, since their masses have not been measured from their orbital
parameters.  Super-Eddington systems like SS~433 and Cyg X--3 are
excluded from this list because neither pulsations or bursts would be
expected even if they contain neutron stars, making identification of
the compact object difficult.  Also excluded from this definition are
systems with a very large variability that spend long periods with
little or no emission, such as GRS~1915+105 and GX~339--4; these are
intermediate between the persistent sources and the x-ray novae, which
have rare, bright, discrete outbursts lasting weeks to months.  Within
the class of five systems thus defined, there is a wide range of
secondaries.  In order of descending mass, \cyg\ has an O9.7Iab (blue
supergiant) companion \citep{Gi86}, \lmo\ a blue giant of type O7III
\citep{Co95}, \lmc\ a massive B3V main sequence star \citep{Wa75}, and
\grs\ and \onee\ are thought to contain low-mass red giants
\citep{Ma98, Ro02, Sm02a}.

\cyg, being the brightest of these systems and the first discovered,
has long been treated as the canonical persistent black-hole binary,
and models of the accretion and x-ray emission processes have often
concentrated on explaining its behavior.  In \cyg, the x-ray
luminosity is higher when the spectrum is soft (dominated by the
thermal component, with an additional power-law component of index
$\sim$--2.4) and lower when the spectrum is hard (power-law dominated,
with index $\sim$--1.7).  There is no time delay; the two quantities
evolve simultaneously \citep[e.g.,][]{Sm02b,Po03}.  A different behavior is
seen in outbursts of x-ray novae: a transition from hard to soft at
the peak of luminosity, followed by the reverse transition at a much
lower level during the decay of the outburst.  This has been
called ``hysteresis'' and has been observed and discussed extensively
in the literature \citep[e.g.,][]{Mi95,Ho01,Ma03}.

In earlier papers \citep{Ma99,Sm02b,Po06} we demonstrated a hysteresis
effect for changes within the hard state in the persistent binaries
\onee\ and \grs.  The result, clear in both sources, is that the
power-law index correlates with the opposite of the derivative of the
photon flux; i.e. the spectrum is softest while the photon flux is
dropping. Further, there is a time-delay:
the power-law index leads the change in photon flux by about 10 dy.
In \citet{Ma99} we only had a single, slow rise and fall in photon
flux to observe in each source; at that time, the pattern of changes
could have been a simple time delay.  In the later papers, there are
both gradual and abrupt drops in photon flux, and we can see that the
characterization of spectral dependence on the derivative is correct:
the spectrum softens much more violently when the count rate drops
more quickly.

Our qualitative explanation for this behavior is discussed in \S 4.1
Briefly, we assumed that a hot corona upscatters nearly all the x-rays
from a thin disk present even in the hard state \citep{Ch97}.  The
coronal and disk accretion flows are fed simultaneously, and when the
mass flow drops it drops first in the corona -- softening the spectrum
-- and, after a viscous delay, in the thin disk, lowering the count
rate \citep{Ch95}.  Adopting a picture so obviously oversimplified was
justified by its ability to explain several different features of the
data, both slow changes over months \citep{Ma99} and two rapid changes
over days: a sudden shutoff of \grs, in which the hard component
vanished first \citep{Sm01}, and a brief, temporary hardening in \onee,
with no change in photon flux, interpreted as a brief ``puff'' of
extra coronal material, too short to have a noticeable influence on
the disk later on \citep{Sm02b}.

\section{Observations}

In March 2005, we began twice-weekly observations of \lmc\ with the
{\it Rossi X-ray Timing Explorer (RXTE)},
as an extension to our original monitoring program of \onee\ and \grs.
The system had been observed many times by {\it RXTE} in the past, but never
with the frequency necessary to systematically observe time evolution
on the expected viscous timescale of days to weeks.  Since \lmc\ is usually in
the soft state, our intention was to see if there was any hysteresis
among the measurable quantities in the soft-state spectrum:
temperature and flux of the thermal component and index and flux of
the power law.  Since \lmc\ does not have a supermassive companion,
and probably accretes via Roche-lobe overflow rather than a wind, we
thought it might show effects at the viscous timescale.  The second
goal of the program, after the search for hysteresis, was to serve as
a trigger for deeper observations of the rare hard state.  There has
not been a transition to the hard state during our monitoring so far.

The {\it RXTE} observations used here were target \#03 of proposals 91105
and 92095, for a total of 139 pointings as of August 2006 (the
campaign is still in progress).  The Proportional Counter Array (PCA)
data were analyzed with version 5.3.1 of FTOOLS including version
11.3.1 of XSPEC \citep{Ar96}.  Since the data were not of high statistical
significance, particularly in the power law component, it was
necessary to minimize the number of free parameters in the fit.  We
began with the simple model of a multicolor disk blackbody
\citep{Mi84} plus a power law.  We froze the equivalent hydrogen
column $n_{\rm H}$ at $3.8 \times 10^{20}$cm$^{-2}$ \citep{Pa03}.
The statistics in the power law component were often not good enough to
simultaneously constrain the index and intensity well.  Since the
power law index of black hole binaries in the soft state often saturates
at a value near 2.5 \citep{Ti05}, we decided to freeze the index at a
value typical for \lmc.

\begin{figure}[h!]
\epsscale{0.6}
\plotone{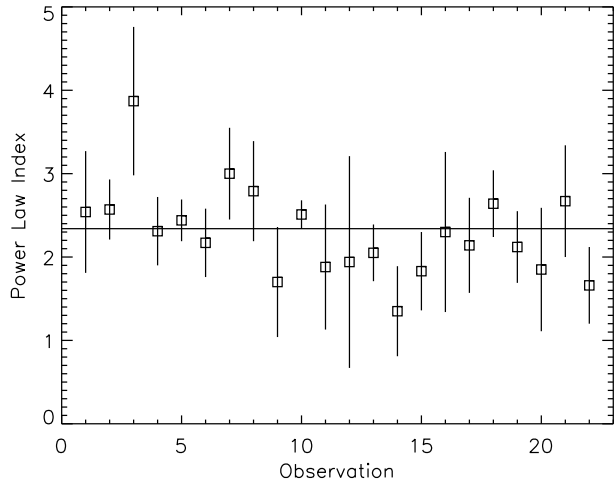}
\caption{
Fitted power-law index from 22 deep pointings to \lmc.  The abscissa
is the chronological order of the observation; since many of the
observations were clustered together, it is presented this way for clarity,
rather than with time on the abscissa.
The best-fit constant index of 2.34 is also shown.
}
\end{figure}

We analyzed 22 deep pointings ($>$ 10 ks) to \lmc\ from earlier in the
{\it RXTE} mission (MJD 50318 to 51267) to see if they had a relatively
narrow distribution of indices and select the best value.  Because the
counting statistics in the bright disk blackbody component are orders
of magnitude better than those in the power-law tail, we found that
tiny discrepancies between the model and the data in the 3--6~keV band
would force the best overall fit to be very poor in the
power-law-dominated part of the spectrum (above 10~keV) \citep{Wi01}: the power-law
indices were clearly much too soft.  When we restricted these fits to
the 6--20~keV range, the model fit much better where the power law
dominated, and gave harder indices.  A similar result was obtained by
adding a 1\% systematic error to the data points and fitting over the
full range (3--20~keV).  The softer indices from our first attempt at
fitting were consistent with those reported by \citet{No01}.
Figure~1 shows the new power-law indices from the deep pointings.
The mean of the distribution (optimally weighted) is 2.34 and the
reduced $\chi^2$ for the hypothesis of a constant value is 0.82.  We
kept the power-law index fixed at this value for the fits to our
monitoring observations.  This gave us three parameters for each fit:
the power-law flux, disk blackbody flux, and disk blackbody temperature.

\section{Results}

We expect a close correlation between the disk blackbody flux and
temperature in the absence of any change in geometry such as a change
in the inner radius or a change in inclination.
This correlation is indeed present, as shown in
Figure~2.  The model shown with the data is the disk blackbody of
\citet{Mi84} integrated from 2--10~keV, as was the fit to each
observation.  It has been normalized to match the data but there are
no other free parameters.  As expected, there is no
hysteresis between these parameters (which would spoil the
correlation); this suggests that there was no major change
in the disk inner radius or inclination angle over the $\sim$1.5~yr of
this campaign.  This comfirms the results of \citet{Eb93} and
\citet{Wi01}, although see \citet{Me00} for complications related to
the interpretation.  

\begin{figure}[h!]
\epsscale{0.6}
\plotone{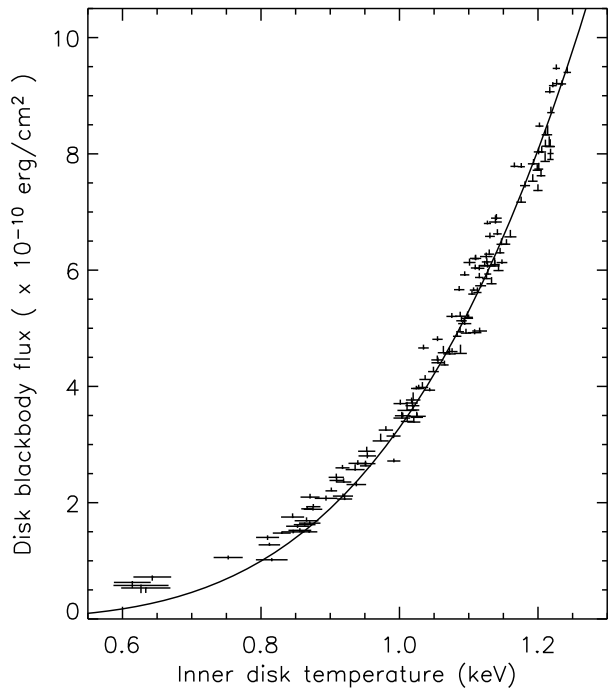}
\caption{
Fitted disk blackbody flux (2--10~keV) as a function of disk blackbody
temperature for the monitoring observations of \lmc\ in {\it RXTE}
cycles 10 and 11.  The model shown for comparison is the expectation
for a disk blackbody with constant geometry, and only the accretion
rate varying.  It has been normalized to match the data but has no
other free parameters. 
}
\end{figure}

Since these parameters are so closely correlated, either can be
equivalently compared to the power-law flux.  Figure~3 shows the
power-law normalization and disk-blackbody flux as a function of time.
A hysteresis effect is immediately obvious: every peak in the
power-law flux is related to a drop in the blackbody flux, and the
size of the peak is proportional to the size of the drop.  Since the
effect is asymmetrical in time and repeated several times, it cannot
be due to any issue of the fitting procedure (e.g.  deviations from a
true disk blackbody distorting the power-law fit), since such effects
would have no sensitivity to the time-ordering of the observations.

If we assume that there is an optically thin corona above an optically
thick, geometrically thin thermal disk, then the photon flux from the
thermal component, $F_{\rm disk}$, depends primarily on the mass
accretion rate \mdotd\ proceeding through the thin disk and
not on its surface density $\Sigma_{\rm disk}$.  The 
accretion rate \mdotd\ is proportional to the $\frac{5}{6}$
power of the observed flux \citep{Fr02}.  In steady state,
there is also a relation between \mdotd\ and $\Sigma_{\rm
disk}$, but we want to retain the possibility of being out of
equilibrium with respect to the viscous time scale.  The flux seen in
the Comptonized power-law tail, $F_{\rm corona}$, should go
approximately as $F_{\rm disk} \Sigma_{\rm corona}$; so we can
approximate $\Sigma_{\rm corona}$ by $F_{\rm disk} / F_{\rm corona}$.
  
\begin{figure}[h!]
\epsscale{1.0}
\plotone{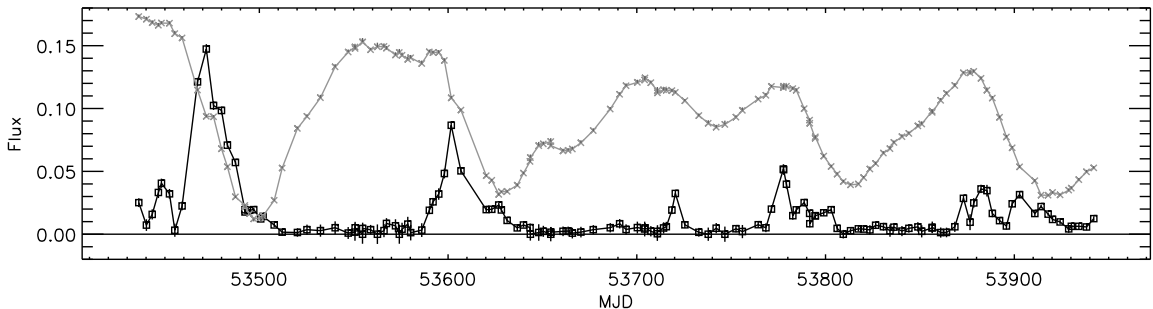}
\caption{
Intensity of the power law (black squares) and disk blackbody (grey
crosses) components of the \lmc\ spectrum versus time.  The power law
component is in units of photons keV$^{-1}$cm$^{-2}$s$^{-1}$ at 1 keV.
The disk blackbody is in units of $10^{-10}$erg cm$^{-2}$s$^{-1}$.
Since we are freezing the power law index at a fixed value (see text),
the energy flux in the power law would have an identical time
dependence.
}
\end{figure}

\begin{figure}[h!]
\epsscale{1.0}
\plotone{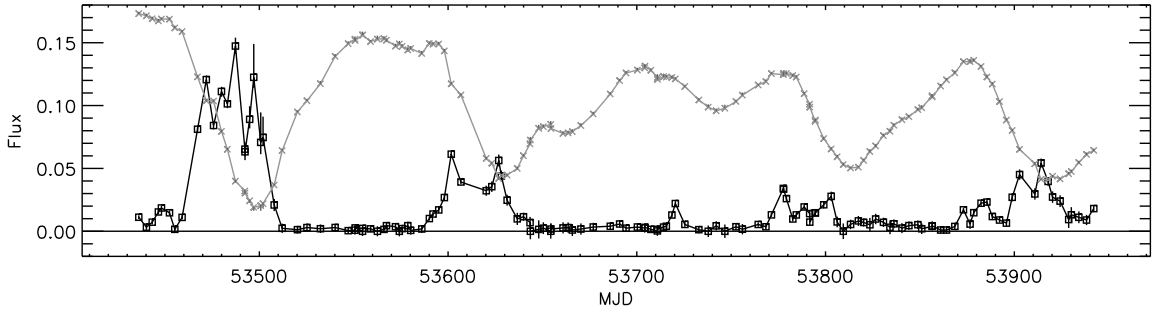}
\caption{
Derived surface density of the hot corona (black squares)
shown with the derived accretion rate in the inner part of the thin
disk (grey crosses) for \lmc,
versus time.  The scaling is arbitrary here, and has been chosen 
to resemble the previous figure.  See text.
}

\end{figure}

Figure~4 shows \mdotd\ and $\Sigma_{\rm corona}$ with these
assumptions.  For a highly sub-Keplerian corona in near free-fall,
\mdotc\ $\sim \Sigma_{\rm corona}$ as was assumed in
\citet{Sm02b}, but the result of Figure~4 is more general if
interpreted as \mdotd\ and $\Sigma_{\rm corona}$ without
this assumption.  We see that the corona turns on and shuts off rather
symmetrically (this would not be apparent when looking at the raw data
of Figure~3).  The corona seems to be present during the decline of
\mdotd\ and shuts off only when \mdotd\ begins
to rise again.

\begin{figure}[h!]
\epsscale{0.6}
\plotone{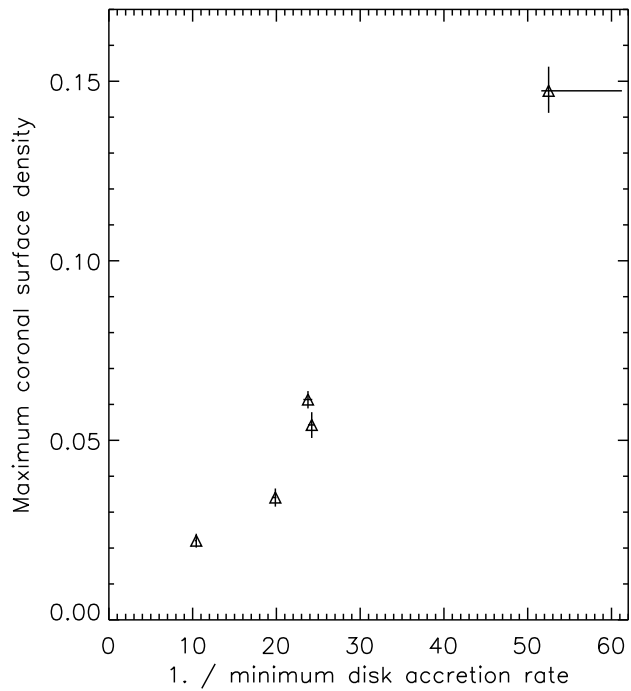}
\caption{
Correlation between the maximum value of the surface density of the
hot corona and the depth of the subsequent minimum in the inner-disk
accretion rate (see text).  Units are arbitrary.
}
\end{figure}

\begin{figure}[h!]
\epsscale{0.6}
\plotone{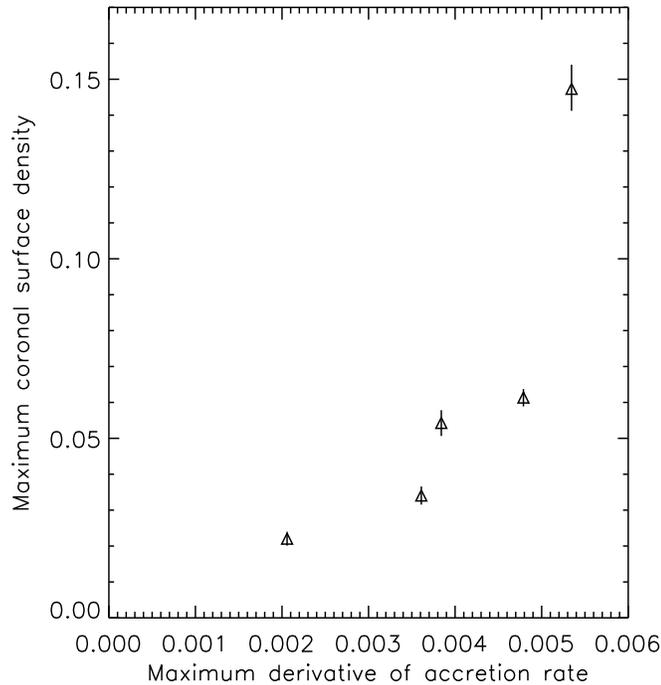}
\caption{
Correlation between the maximum value of the surface density of the
hot corona and the maximum rate of decline in the associated drop
in the inner-disk accretion rate (see text).  Units are arbitrary.
}
\end{figure}

The size of the peak in $\Sigma_{\rm corona}$ seems to increase as
the following dip in \mdotd\ gets deeper.  To quantify this,
we identify five discrete episodes of power-law emission,
peaking at MJD 53487, 53602, 53720, 53778, and 53914.  The next minima
in the curve of \mdotd\ appear at MJD 53497, 53627, 53742, 53813,
and 53918, an average of 19 days later.  Figure~5 shows the maxima in 
$\Sigma_{\rm corona}$ as a function of the reciprocal of the corresponding
minimum \mdotd.  Figure~6 shows the same maxima of coronal density plotted, instead,
against the maximum steepness of the associated decline (in arbitrary units).
More data are needed to see which correlation better characterizes the situation.

\section{Discussion}

\subsection{Interpretation in a two-flow picture}

\citet{Ch95} originated the concept we used in \citet{Sm02b}, of two
separate accretion flows, one Keplerian and one sub-Keplerian,
co-existing at all radii, that could be fed separately.  Hysteresis
effects come about due to the viscous delay in the Keplerian disk; the
changes in accretion rate propagate to the central regions, where
x-rays are created and upscattered, first in the sub-Keplerian flow
and only later in the Keplerian flow.  We have used the adjective
``independent'' to indicate that in this picture the state of the
corona is not entirely and uniquely determined by the instantaneous
state of the thin disk.  It has been pointed out, rightly, that the
flows are not literally independent of each other, since they will
interact via the exchange of energy by radiation and conduction and
the exchange of matter by condensation and evaporation \citep{Me07}.

In \citet{Ma99} and \citet{Sm02b} we discussed the hysteresis we found
in the hard-state persistent systems \onee\ and \grs\ in terms of this
two-flow picture.  The spectrum, consisting entirely of the
inverse-Comptonized hard power law, was softest just before and during
the decline of the count rate.  We postulated that the inner thin disk
continued to exist beneath the Comptonizing corona; thus the total
number of photons observed (all in the power-law component) was taken
as equivalent to the number of thermal seed x-rays generated in the
standard thin disk.  Recent spectral analysis of the hard state in
\gx\ and \cyg\ \citep{Mi06} lends strong support to this picture of a
hard state with no truncation of the thin disk, at least in some
cases.

We interpreted the hysteresis between the observed parameters (power
law index and count rate) as indicating that the material in the
Comptonizing corona drops just before the underlying thin disk begins to
decline.  This might be expected from a drop in a ``master'' accretion
rate that feeds both components in a fixed ratio.  The hysteresis is
the natural effect of the sub-Keplerian corona responding faster than
the Keplerian disk \citep{Ch95}.  The period of soft emission in which
the sub-Keplerian material has accreted away and the thin disk is still 
slowly dropping, we call a ``dynamical soft state,'' with the transition 
of \grs\ in 2001 as the canonical example \citep{Sm01}.  The contrasting
``static'' soft state would occur when
the accretion rate is high enough that the corona is unstable to collapse
by bremsstrahlung cooling.  This has never been observed in \onee\ or \grs,
which stay well below 10\% of the Eddington luminosity for a black hole.

Since the thermal blackbody emission is clearly visible throughout all
our data from \lmc, and since it also seems that the disk geometry
isn't changing (Figure~2), the interpretation here is more direct than
for the hard-state systems: the directly-measured thermal flux is now
the proxy for the accretion rate in the inner disk, while the
percentage of the total flux that is in the power-law tail is the
proxy for the amount of material in the inner corona (see above).  The
power-law flux, rather than the power-law index, is used since the
corona is now very optically thin rather than marginally optically
thick as it is in the hard state.

Figure~4 then suggests that the coronal density is high while the thin
disk is decreasing, and at its minimum.  If our interpretation of the
hard-state systems was correct, then, the new soft-state data from
\lmc\ are showing a very different effect in the underlying accretion
flows: rather than the ``master'' accretion rate rising and falling
together for both flows, it may instead be temporarily diverted so as
to favor the corona over the thin disk, ``starving'' the latter while
the former is high.

\subsection{Unification with other types of hysteresis}

Previously, hysteresis has mostly been studied in the case of
transient outbursts (called x-ray novae or soft x-ray transients) from
black hole \citep{Mi95} or neutron star \citep{Ma03} binaries with
low-mass companions.  Are we now faced with three physically unrelated
types of hysteresis, for hard persistent systems, soft persistent
systems, and transients, all with similar timescales of days to weeks?
Perhaps some unification is possible.
 
The hysteresis seen in the persistent hard-state systems, associated
with either slight softenings or full transitions to the dynamical soft
state, may operate the same way as the hard-to-soft transitions at the
peak of x-ray novae.  A recent study of this transition in many
transients \citep{Gi06} finds that they can be divided into two
classes by the rate of the transition -- those that make the
transition quickly tend to do so at lower luminosity, and the
transitions at highest luminosities tend to proceed slowly.
\citet{Gi06} refer to these as ``dark/fast'' and ``bright/slow''
transitions.  They note that the dark/fast type can occur over a range
of luminosities in the same binary (for example, \gx).  Comparison to
our results for hard-to-soft transitions in \onee\ and \grs\ suggests
that the dark/fast transitions can be identified with our dynamical soft
state, and the bright/slow transitions with what we called the static
soft state.

The increases in the power-law tail in the soft state of \lmc, interpreted
as partial transitions toward the hard state, may be related to the
soft-to-hard transitions at the end of x-ray nova outbursts \citep{Ka04}.
\citet{Ma03a} studied the soft-to-hard transitions in a large sample of
black-hole and neutron-star transients, and found that they generally
occur between 1\% and 4\% of Eddington luminosity.  The lowest
luminosities in Figure~2, corresponding to the periods with the hardest
overall spectrum, occur around 1$\times 10^{-10}$erg cm$^{-2}$s$^{-1}$.
Assuming a distance of 50~kpc and a black-hole mass of 5 solar masses
for \lmc, this corresponds to about 1\% of Eddington luminosity,
suggesting that these partial transitions -- as well as the complete
ones seen by \citet{Wi01} -- could be analogous to the transitions
in the tails of the transients.

\subsection{Other models of hysteresis}

A number of other physical models have been presented to explain
the hysteresis effect in transients.

\citet{Wa96} described the effect of radiation drag on the coronal
flow.  In this picture, hot coronal gas is affected by two radiation
fields: strong radiation from the compact object (in our case,
substitute the innermost portion of the disk) and local radiation from
the nearby part of the thin disk.  The former tends to cause the
coronal gas to fall in and accrete, while the latter tends to keep it
in place.  As changes in accretion rate propagate through the thin
disk, it will go through states with a different ratio of luminosities
in the inner and outer parts.  When the inner disk is more luminous
than the outer, most of the corona will flow inwards while it
evaporates, while in the opposite case there will be more corona
remaining above the disk at all radii.  Here, although the coronal
density cannot be determined entirely by the local conditions in the
thin disk, it is determined entirely by the state of the
whole thin disk, which is out of viscous equilibrium.  The radiation
drag picture can still show hysteresis even without independently
specifying input accretion rates for the two flows, and thus has the
potential of explaining hysteresis phenomena with fewer free
parameters.

\citet{Me05} and \citet{Li05} discuss the effect of irradiation on
heating and cooling of the corona, rather than drag.  The x-rays from
the innermost part of the flow affect the corona further out.  In
the hard state, the hard radiation heats the corona at larger radii,
encouraging it to remain in the hard state during the rise of a
transient.  In the soft state, the soft radiation cools the corona as
it tries to reform during the outburst decay; the soft state thus
helps to perpetuate itself down to lower luminosities.

\citet{Ma06} describe a mechanism in which magnetic pressure can
forestall the bremsstrahlung-cooling instability by which a sub-Keplerian,
optically thick flow collapses into an optically thick disk.  They
thus explain why the hard state can be maintained to
high luminosities in the rise phase of a transient.  Our new results
for \lmc\ probably require another explanation, however, since the
system never leaves the soft state.

Most recently, \citet{Me07} discussed the hysteresis in transient
outbursts in terms of evaporation and condensation of the inner part
of the thin disk.  In this picture, the extension of the soft state to
low luminosities in transients during the decline of the outburst is
due to re-condensation to a thin disk at small radii, even when the
flow is purely advection-dominated at intermediate radii.
\citet{Me07} suggest that the reason \cyg\ shows no hysteresis is that
the innermost recondensation disk never entirely disappears, even in
the hard state, since the accretion rate never gets very small.  Our
current result poses a challenge to this model: from Figure~2, it is
clear that the inner disk exists down to the last stable orbit
throughout our data set, yet there is a strongly hysteretic behavior
still present.

\citet{Ma03} also suggested that the lack of hysteresis in \cyg\ is
due to the relatively small changes in luminosity accompanying state
transitions.  However, our earliest results on hard-state hysteresis
\citep{Ma99} seem to be at odds with that conclusion, since on
those occasions \onee\ and \grs\ showed hysteresis during very minor
changes in luminosity (less than a factor of two).  We continue to
suggest that the size and viscous timescale of the disk, expected to
be very small for a wind accretor, is the relevant parameter that
distinguishes \cyg\ from the other systems.

\acknowledgments

This work was supported by NASA grants
NNG04GP41G and NNG05GM71G.  The authors thank Nathan
Bezayiff for important contributions to the software used in
the data analysis, and Enrico Ramirez-Ruiz and Douglas Lin
for valuable discussions.

\end{sloppypar}
\end{document}